\documentclass{aa}

\usepackage{graphicx}
\begin{document}
   \title{$^{13}$C isotope effects on infrared bands of quenched carbonaceous 
composite (QCC)}

   \author{Setsuko Wada           \inst{1}
          \and  Takashi Onaka  \inst{2}
          \and  Issei Yamamura  \inst{3}
          \and  Yoshitada Murata  \inst{1}
          \and  Alan T. Tokunaga  \inst{4}
          }

   \offprints{S. Wada, \email{wada@e-one.uec.ac.jp}}

   \institute{Department of Applied Physics and Chemistry,
              The University of Electro-Communications, 
              Chofugaoka, Chofu, Tokyo 182-8585, Japan
   \and       Department of Astronomy, The University of Tokyo, 
              7-3-1 Hongo, Bunkyo-ku, Tokyo 113-0033, Japan
   \and       The Institute of Space and Astronautical Science (ISAS), 
              Yoshino-dai 3-1-1, Sagamihara, Kanagawa 229-8510, Japan
   \and       Institute for Astronomy, University of Hawaii,
              2680 Woodlawn Dr., Honolulu, HI 96822, U.S.A.     
  }            

   \date{Received 6 May 2003 / Accepted 2 June 2003}

   \abstract{
We investigate carbon isotope effects on the infrared bands of a laboratory analogue
of carbonaceous dust, the quenched carbonaceous composite (QCC), synthesized
from a plasma gas of methane with various \element[][12]{C}/\element[][13]{C}
ratios. Peak shifts to longer wavelengths 
due to the substitution of \element[][12]{C}
by \element[][13]{C} are clearly observed
in several absorption bands.  The shifts are almost linearly proportional to
the \element[][13]{C} fraction.  New features associated
with \element[][13]{C} are not
seen, indicating that the infrared bands in the QCC are not very
localized vibration modes but come from vibrations associated with rather large
carbon structures.
An appreciable peak shift ($\Delta \lambda \sim $
0.23--0.26\,$\mu$m per \element[][13]{C} 
fraction) 
is detected in the 6.2\,$\mu$m band, which is
attributed to a carbon-carbon vibration. 
A peak shift ($\Delta \lambda \sim $
0.16--0.18\,$\mu$m per \element[][13]{C} fraction) in an
out-of-plane bending mode of aromatic
C$-$H at 11.4\,$\mu$m is also observed, while only a small shift 
($\Delta \lambda <$ 0.015\,$\mu$m per \element[][13]{C} fraction) is
detected in the 3.3\,$\mu$m band, which arises from a 
C$-$H stretching mode. 
The present experiment suggests that  
peak shifts in the unidentified infrared (UIR) bands,
particularly in the 6.2\,$\mu$m band, should be 
detectable in celestial objects with 
low \element[][12]{C}/\element[][13]{C} ratios ($< 10$).  
The isotopic shifts seen in the QCC are discussed in relation
to the variations in the UIR band peaks observed 
in post-asymptotic giant branch stars
and planetary nebulae.  
The observed peak shift pattern of the UIR bands
is qualitatively in agreement with the 
isotopic shifts in the QCC except for the 7.7\,$\mu$m band
complex although the observed shifts in 
the UIR bands are larger than
those inferred from derived isotope ratios for individual
objects.  The poor quantitative agreement may be attributed partly to  
large uncertainties in the derived \element[][12]{C}/\element[][13]{C},
to possible spatial variations of the isotope abundance
within the object, and to combinations of other effects, such as
hetero-atom substitutions.  The present investigation suggests that part of
the observed variations in the UIR band peaks may come from the isotopic 
effects.

   }

\titlerunning{Carbon Isotope Effects on IR Bands}
\authorrunning{Wada, Onaka, Yamamura et al.}

\maketitle

%
\section{Introduction}

A wide range of the \element[][12]{C}/\element[][13]{C} ratio has been reported 
in various celestial
objects.  The ratio changes due to nucleosynthesis 
and mixing in the interior of stars.  During the first dredge-up on the red
giant branch (RGB) the convective envelope reaches regions abundant in
\element[][13]{C} that was
processed from \element[][12]{C} and the ratio decreases. 
Observations of RGB stars often
show the \element[][12]{C}/\element[][13]{C} ratio
of 5--20, even lower than theoretically
predicted, suggesting the presence of extra-mixing below the 
convective envelope (e.g. Gilroy \cite{gil89}).
In the third dredge-up during the asymptotic giant branch (AGB) phase, an 
increase in
\element[][12]{C}/\element[][13]{C} is generally expected, but the ratio 
could also decrease
due to cool bottom processing for low-mass stars (Wasserburg et at. 
\cite{was95}; Nollett et al. \cite{nol03})
or to hot bottom burning for more massive stars (Frost et al. \cite{fro98}).
Observations of five carbon-rich circumstellar envelopes indicate the
ratio of 30--65 (Kahane et al. \cite{kah92}), while ten carbon stars are shown
to have the ratio in the range 12--60 in their circumstellar envelopes
(Greaves \& Holland \cite{gre97}).
Some carbon stars show the ratio of \element[][12]{C}/\element[][13]{C} as low 
as 3
(Lambert et al. \cite{lam86}; Ohnaka \& Tsuji \cite{ohn96}; Sch\"oier 
\& Olofsson \cite{sch00}). 
The \element[][12]{C}/\element[][13]{C} ratio in post-AGB stars and planetary 
nebulae (PNe)
reflects the cumulative effects of different mixing and nuclear processing
events during the entire evolution of their progenitors.  Lower limits
of the ratio of 3 to 10 have been obtained for several objects in
the post-AGB phase (Palla et al. \cite{pal00}; Greaves \& Holland 
\cite{gre97}).  Recently Josselin \& L\'ebre (\cite{jos01}) estimated
an upper limit of \element[][12]{C}/\element[][13]{C} of 5 for the
post-AGB candidate, HD179821,  whereas a relatively
large ratio of $72 \pm 26$ is reported for another post-AGB star, 
HD56126 (Bakker \& Lambert
\cite{bak98}).
Clegg et al. (\cite{cle97}) found low ratios of 15 and 21 
in two PNe.  Further low values of the ratio of 2--30 have been reported in 
recent
studies of several PNe 
(Palla et al. \cite{pal00}; Balser et al. \cite{bal02}; Josselin \& Bachiller
\cite{jos03}), suggesting that some stars undergo non-standard processing
in the stellar interior and a low \element[][12]{C}/\element[][13]{C}
can be expected during the late stage of their evolution.  The solar 
system value is 89 (Anders \& Grevesse \cite{and89}).

The \element[][12]{C}/\element[][13]{C} ratio also provides key information on 
the
chemical evolution in the Galaxy (for a review, Wilson \cite{wil99}).
Observations of molecules and solid CO$_2$ in interstellar medium
indicate that the ratio ranges from 10--100 and increases with the 
Galactocentric
distance.  The \element[][12]{C}/\element[][13]{C} ratio is suggested to be
about 10--20 in the Galactic center region
(Wilson \cite{wil99}; Boogert et al. \cite{boo00};
Savage et al. \cite{sav02}).  The Galactic
gradient is thought to be built by nucleosynthesis
of the Galactic chemical evolution
and the suggested ratio of 10--20 in the Galactic center
indicates the presence of significant
stellar sources of \element[][13]{C}.  
Interstellar graphite spherules in the Murchison meteorite show
a range of the ratio of 7--1330 (Bernatowicz et al. \cite{ber91}).
Some presolar SiC grains show very low
\element[][12]{C}/\element[][13]{C} ratios of less than 10 and they are 
thought to originate
from very \element[][13]{C}-rich stars in the AGB phase (Amari et al. 
\cite{ama01}).
These observations suggest that low \element[][12]{C}/\element[][13]{C} ratio 
environments
are not uncommon in objects in the AGB, post-AGB, and PN phases as well as in
some interstellar medium.  Carbon-bearing species formed
in these environments could thus
show non-negligible carbon isotopic effects in their spectrum.

A set of emission bands at 3.3, 6.2, 7.6--7.8, 8.6, and 11.2\,$\mu$m have
been observed in various celestial objects and are called
the unidentified infrared (UIR) bands.  Fainter companion bands
are also sometimes seen.  The exact nature of the carriers has not yet
been understood completely, but it is generally believed that the 
emitters or emitting atomic groups containing polycyclic aromatic 
hydrocarbons (PAH) or PAH-like atomic groups of carbonaceous materials,
including such as nanodiamond grains, are responsible for the UIR bands
(L\'eger \& Puget \cite{leg84}; Allamandola et al. \cite{all85}; Sakata et al. 
\cite{sak84};  Papoular et al. \cite{pap89}; Arnoult et al. \cite{arn00};
Jones \& d'Hendecourt \cite{jon00}).
Alternatively, Holmid (\cite{hol00}) has recently proposed
de-excitation of Rydberg matters as possible carriers.  
The UIR bands have been observed in 
a wide range of objects, including \ion{H}{ii} regions, reflection nebulae,
post-AGB stars, and PNe (for a review, see Tokunaga \cite{tok97}).  
They have also been commonly seen in the diffuse Galactic
emission (Tanaka et al. \cite{tan96};
Onaka et al. \cite{ona96}; Mattila et al. \cite{mat96}; Kahanp\"a\"a et al.
\cite{kah03}) as
well as in external galaxies (e.g. Mattila et al. \cite{mat99};
Helou et al. \cite{hel00}; Reach et al. \cite{rea00}; Lu et al. \cite{lu03}),
indicating that the carriers are a common member of interstellar medium and
present in various environments.
Carbon-rich objects in the evolutionary stage from post-AGBs to PNe often show 
the emission bands and thus isotopic effects should be detectable if they
arise from carbonaceous materials of low \element[][12]{C}/\element[][13]{C} 
ratios.
Simple calculations of a \element[][13]{C}-benzene molecule suggest
that the peak shift can be as much as 0.15\,$\mu$m for the C$-$C 
stretching mode (Appendix \ref{cal}).

Observations of the Infrared Space Observatory (ISO; Kessler et al. \cite{kes96}) 
have provided a large
database of the UIR band spectra in various objects (e.g. 
Beintema et al. \cite{bei96}; Molster et al. \cite{mol96}; 
Verstraete et al. \cite{ver96, ver01}; Boulanger
\cite{bou98}; Cesarsky et al. \cite{ces00a, ces00b}; 
Uchida et al. \cite{uch98, uch00}; Moutou et al. \cite{mou00}; 
Honey et al. \cite{hon01}). 
Recently Peeters et al. (\cite{pee02}) have investigated in detail 
the 6--9\,$\mu$m spectra of 57 
sources taken by the Short Wavelength Spectrometer (SWS; de Graauw et al. 
\cite{degr96}) on board the ISO and
found that the 6.2, 7.7, and 8.6 $\mu$m UIR bands show appreciable
variations particularly for post-AGB stars and PNe.  On the other hand, 
the variations in the 11.2\,$\mu$m band are relatively modest and those
in the 3.3\,$\mu$m are less pronounced (Tokunaga et al. \cite{tok91};
Roche et al. \cite{roc91};
Hony et al. \cite{hon01}; van Diedenhoven et al. \cite{van03}).
These variations can be interpreted in part by nitrogen substitutions in
PAHs and anharmonicity, but not all of the observed aspects of
the UIR bands have yet been fully understood (Verstraete et al. \cite{ver01};
Pech et al. \cite{pec02}; Peeters et al. \cite{pee02}).
Part of the observed variations could also originate from  
isotopic effects of the UIR band carriers since the objects that show
the variations are mostly post-AGB stars and PNe, in which small 
\element[][12]{C}/\element[][13]{C} ratios can be expected.

In the present paper we investigate
isotopic effects on the UIR bands experimentally.  We synthesize a 
laboratory analogue of carbonaceous dust, the
quenched carbonaceous composite (QCC; Sakata et al. \cite{sak84}),
with various \element[][12]{C}/\element[][13]{C} ratios from the starting gas 
of a mixture of 
\element[][12]{C}H$_4$ and 
\element[][13]{C}H$_4$.  The QCC shows infrared 
bands similar to the UIR bands and shifts in the band peaks 
due to the \element[][13]{C} substitution
are clearly detected.  In Sect.2 we describe the experimental procedure.
The results are shown in Sect. 3 and discussed in comparison 
with observations in Sect. 4.  A summary is given in Sect. 5.  

\section{Experimental}
The experimental procedure for synthesizing the QCC is described in detail 
in Sakata et al. (\cite{sak84}).  
Methane (CH$_4$), the source gas of the QCC, is decomposed by the imposed
microwave radiation and becomes a plasma. 
Carbonaceous condensates are formed in the injection beam of the
plasma by quenching of the gas. 
Typically two types of the QCC are formed. One is a brown-black material
(hereafter called dark-CC), which is collected on a substrate in the main
injection beam.  It has been shown to consist of a coagulation of 
carbon-onion-like 
particles (Wada et al. \cite{wad99}). The other 
is a yellow-brownish material (hereafter filmy-QCC), deposited 
on the surrounding
region of the injected plasma gas (Sakata et al. \cite{sak87}).  It is a
material rich 
in organic molecules, such as PAHs.  We prepare the starting gas of CH$_4$
with the \element[][13]{C} fraction of 1\%, 11\%, 
25\%, 45\%, 65\% and 99\% by mixing 99\% \element[][13]{C} methane with the 
natural 
isotope abundance (1\% \element[][13]{C}) methane.
We assume that the \element[][13]{C} fraction in the QCC is equal to that 
of the 
starting gas in the present analysis because the reactions in the present
experiments take place at high temperatures.

The QCC is collected either on a KBr (for filmy-QCC) or on a BaF$_2$ crystal 
(for dark-QCC)
and the absorption spectra are measured at room temperature. 
The spectra are taken by a Perkin-Elmer 2000-FTIR spectrometer with the 
resolution of 4 cm$^{-1}$.  The spectra of the dark-QCC are measured after 
washing it with a small amount of acetone and removing organic molecules in the 
dark-QCC in a similar manner to previous experiments (Sakata et al. 
\cite{sak84}).

\section{Results}
The isotopic substitution gives rise to a shift in the wavelength of   
vibration modes due to mass effects.
With the replacement of \element[][12]{C} with \element[][13]{C}, the infrared 
band is expected to be 
shifted to longer wavelengths.

\subsection{Spectra of filmy-QCC}

The filmy-QCC is a coagulation of many kinds of organic molecules.
Mass analysis of the gases evaporating from the filmy-QCC shows that they 
contain various kinds of PAHs.  The filmy-QCC 
also emits red fluorescence under ultraviolet 
irradiation (Sakata et al. \cite{sak92}).

\begin{figure}
   \includegraphics[width=\hsize]{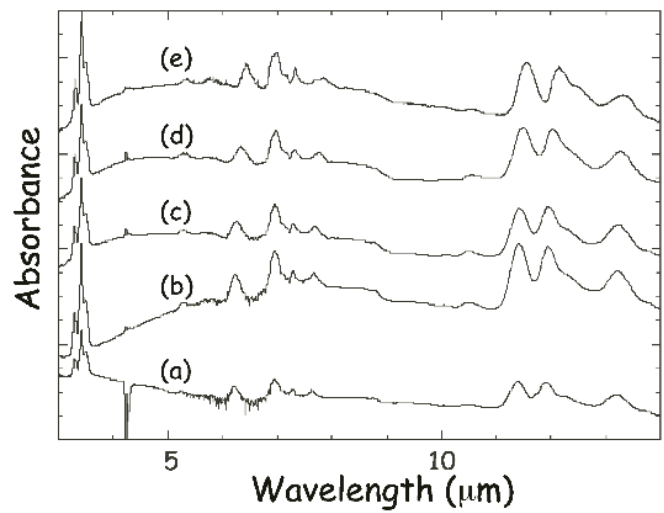}
     \caption{Absorption spectra of filmy-QCC with various \element[][13]{C}
              fractions.
              The \element[][13]{C} fractions are (a) 1\%, (b) 11\%, (c) 
25\%, 
              (d) 65\%, and (e) 99\%.  Each curve is shifted arbitrarily in 
the 
              vertical direction.}
     \label{fig1}
  \end{figure}

\begin{figure}
   \includegraphics[width=\hsize]{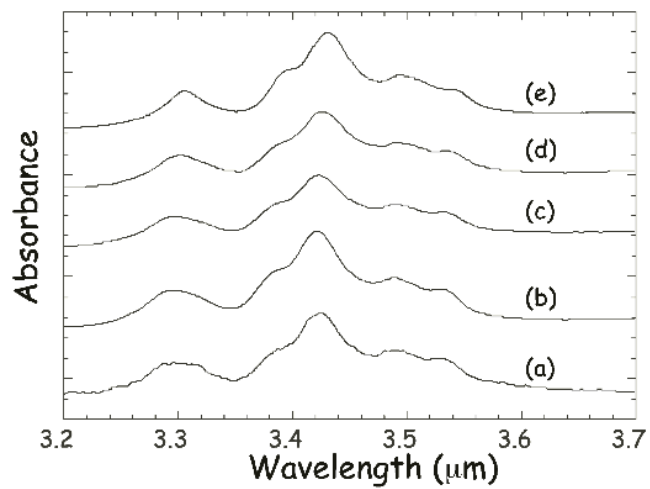}
    \caption{Spectra of the filmy-QCC in the 3\,$\mu$m region.
The \element[][13]{C} fractions are (a) 1\%, (b) 11\%, (c) 25\%, (d) 
65\%, 
and 
(e) 99\%.  Each curve is shifted arbitrarily in the 
              vertical direction.}
\label{fig2}
\end{figure}

The spectra of the filmy-QCC with various \element[][13]{C} fractions
are shown in Fig.~\ref{fig1}.
They show several absorption features similar to the UIR bands observed in 
celestial objects (Sakata et al. \cite{sak84, sak87, sak90}).
For the filmy-QCC with 1\% \element[][13]{C} 
the bands of the C$-$H stretching characteristic group appear at
3.29, 3.42, 3.51, and 3.53\,$\mu$m (see also Fig.~\ref{fig2}). An aromatic 
C$=$C 
bond vibration 
appears at 6.2$\,\mu$m, and aromatic C$-$H out-of-plane bending modes are 
observed at 11.4, 11.9, and 13.2\,$\mu$m.  A weak feature 
also appears clearly at 
7.6\,$\mu$m in addition to the bands at 7.0 and 7.3\,$\mu$m.  A very broad 
band 
is seen around 8.6\,$\mu$m, but its
peak wavelength cannot be determined accurately because of the weak and broad
feature.  
All the features observed in the samples of various \element[][13]{C} 
fractions
have corresponding features in the 1\% \element[][13]{C} sample and the peak
wavelengths are shifted to longer wavelengths with the \element[][13]{C} 
fraction. 
No new absorption band associated with \element[][13]{C} appears.

A large 
shift ($\Delta\lambda \sim 0.23$\,$\mu$m between 1\% to 99\% 
\element[][13]{C})
is observed in the 6.2\,$\mu$m band.
The C$-$C mode involves at least two carbon atoms and makes the shift large.
The shifts in the peaks arising from C$-$H out-of-plane bending modes are
also seen.  The 11.39\,$\mu$m peak at 1\% \element[][13]{C} shifts to 
11.55\,$\mu$m
at 99\% \element[][13]{C}.
On the other hand, the shift in the 3.29\,$\mu$m band of aromatic 
C$-$H stretching is quite small ($\Delta\lambda \sim 0.014$\,$\mu$m).  
Fig.~\ref{fig2} shows the spectra expanded
in the 3\,$\mu$m region.  It is noticeable that the profile
of the 3.3\,$\mu$m band becomes narrower with the \element[][13]{C} 
fraction.
The cause of this change is unknown at present.  A 
small shift is also observed in the 3.4\,$\mu$m band, which is attributed 
to an asymmetric vibration of methylene (C$-$H$_2$).

The peak wavelength of the small feature at 7.6\,$\mu$m
is also shifted to longer wavelengths with the
\element[][13]{C} fraction
($\Delta\lambda \sim 0.22$\,$\mu$m). The large shift indicates 
that the 
vibration involves more than one carbon atoms, such as C$-$C. 
The broad bump at 8.6\,$\mu$m does not shift clearly.
Sakata et al. (\cite{sak87}) assigned the 7.6 and 8.6\,$\mu$m peaks 
of the filmy-QCC to   
vibration modes of a kind of the ketone bond  C$=$C$-$C$=$O.
The 7.6\,$\mu$m peak is always observed in the filmy-QCC spectrum although 
the relative strength changes slightly with individual runs. 
The exposure to air does not increase the strength of the 7.6\,$\mu$m band.
The present experiment thus suggests that the 7.6\,$\mu$m component is a 
molecular product of 
the plasma gas and is not directly related to oxidation.

\subsection{Spectra of dark-QCC}

\begin{figure}
   \includegraphics[width=\hsize]{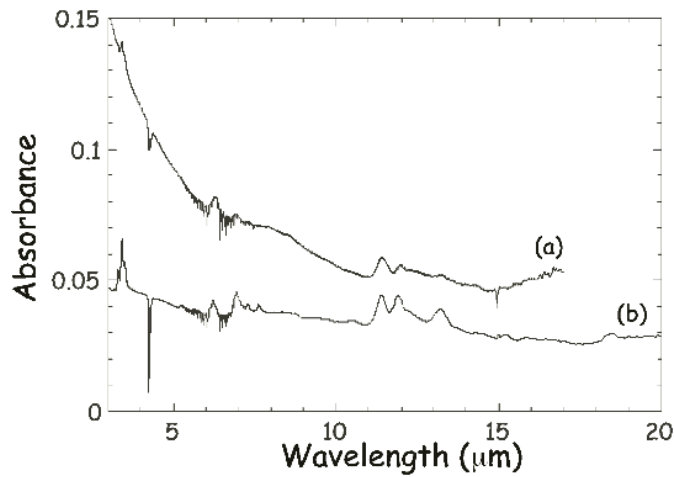}
    \caption{Absorption spectra of (a) dark-QCC  and (b) filmy-QCC 
    with the natural 
           isotopic fraction (\element[][13]{C} = 1.1\%).}
    \label{fig3}
 \end{figure}

High-resolution electron-microscopy reveals that the dark-QCC is comprised
of a coagulation of onion-like particles of the diameter of
10--15\,nm (Wada et al. \cite{wad99}). 
It is insoluble in acetone.
Even after washing with acetone, the dark-QCC still contains some
organic molecules.  Mass spectroscopy 
indicates that they are mainly composed of compact PAHs.
The dark-QCC adsorbs organic molecules that are formed together and also
contains hydrogen atoms in its structure. 
C$-$H bonds in both components will give rise to vibration modes 
in the infrared.
Fig.~\ref{fig3} shows the absorption spectra of the dark-QCC with 1\%
\element[][13]{C} fraction together with the filmy-QCC spectrum for 
comparison.
The dark-QCC shows a strong continuum component, which decreases 
with the wavelength. 
Small peaks are observed at 3.3, 3.42, 5.8,
6.3, 7.0, 7.3, 8.2, 11.4, 12.0, 
and 13.2\,$\mu$m.
The 3.3, 3.42, 11.4, 12.0, and 13.2\,$\mu$m bands are attributed to CH 
vibrations, 
and the 6.3\,$\mu$m band is ascribed to an aromatic C$=$C bond vibration. The 
broad 8.2\,$\mu$m band becomes strong when it is exposed to air.   
An increase of the 5.8\,$\mu$m band is also seen during 
the exposure to air, suggesting 
that both of them are associated with C$-$O$-$ and C$=$O bonds formed by 
oxidation.

\begin{figure}
   \includegraphics[width=\hsize]{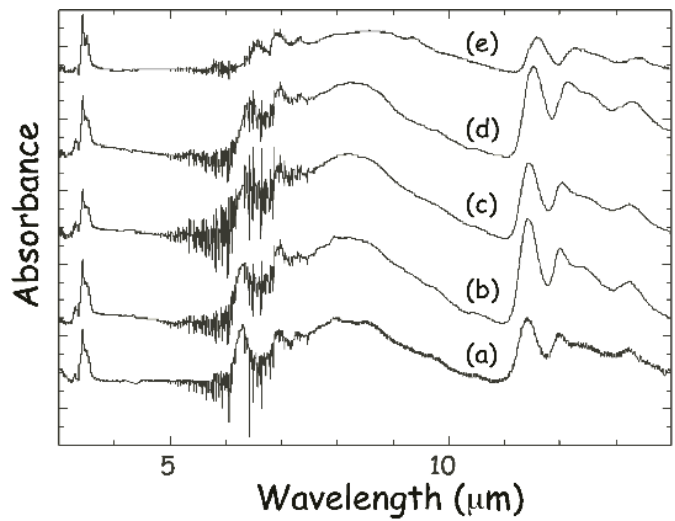}
\caption{Absorption spectra of the dark-QCC with various \element[][13]{C} 
fractions with the continuum subtracted.
The \element[][13]{C} fractions are 1\% (a), 11\% (b), 25 \% (c), 65\% 
(d),
and 99\% (e).  Each curve is shifted arbitrarily in the vertical direction.}
\label{fig4}
\end{figure}

\begin{figure}
   \includegraphics[width=\hsize]{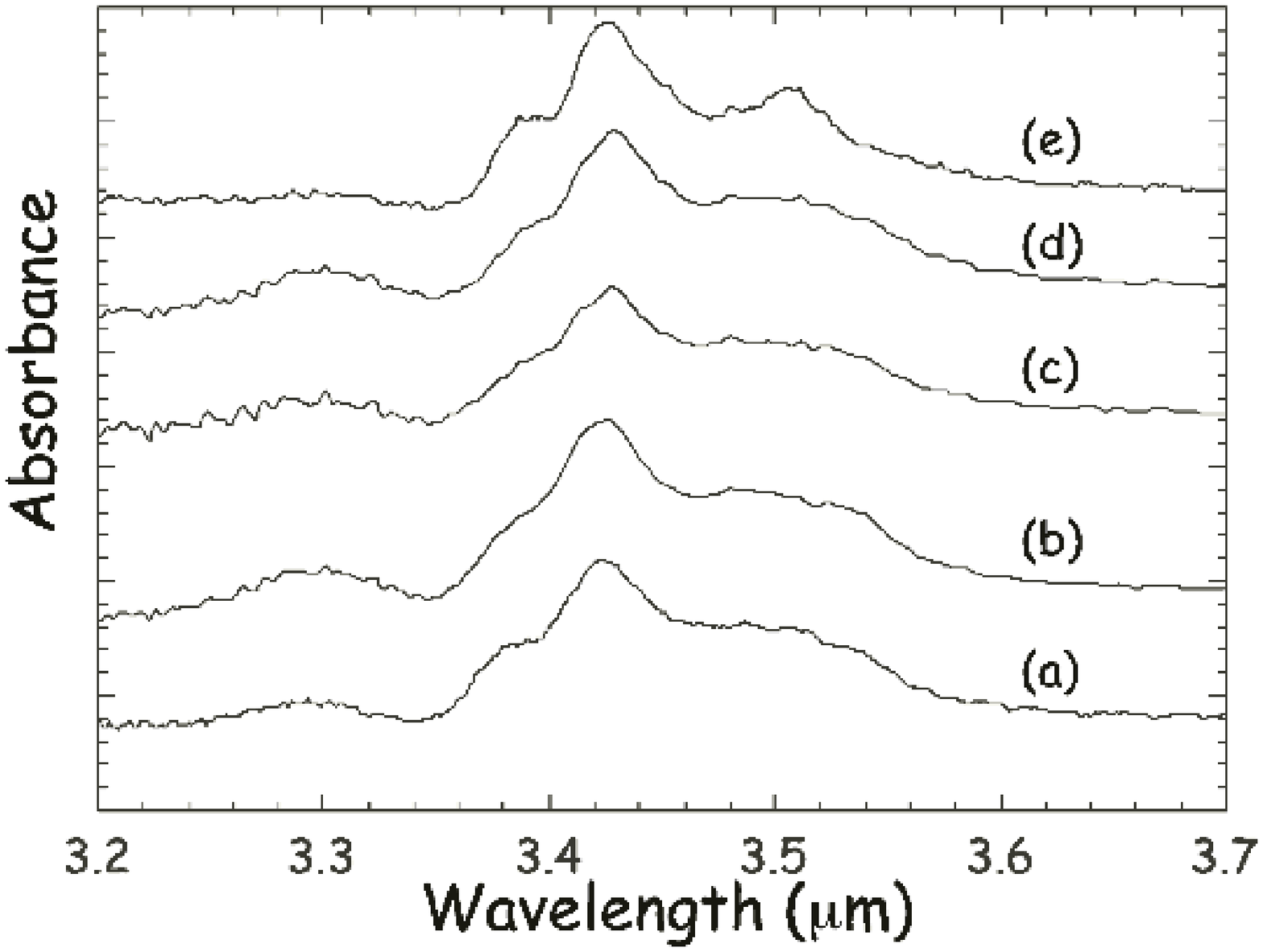}
\caption{Spectra of the dark-QCC in the 3\,$\mu$m region.
The \element[][13]{C} fractions are (a) 1\%, (b) 11\%, (c) 25\%, (d) 
65\%, 
and 
(e) 99\%.  Each curve is shifted arbitrarily in the vertical direction.}
\label{fig5}
\end{figure}

Fig.~\ref{fig4} shows the absorption spectra of the dark-QCC with
various \element[][13]{C} fractions.  The continua were fitted with
cubic spline functions and have been subtracted to
show the absorption features clearly.  As in the filmy-QCC spectra,
peak shifts with the \element[][13]{C} fraction are clearly observed 
and
new features without corresponding ones in the 1\% \element[][13]{C} sample 
are
not seen in large \element[][13]{C} fraction samples.  A large shift 
($\Delta\lambda \sim 0.26$\,$\mu$m) is again observed for the 6.3\,$\mu$m 
band.
The peak position of the 8.2\,$\mu$m band cannot be measured
accurately
because of the wide profile, but the peak is clearly shifted to
longer wavelengths with the \element[][13]{C} fraction. 
The band width also seems to change. 
Particularly for the 99\% \element[][13]{C} sample
the 8.2\,$\mu$m feature becomes very broad.  
The 3.3\,$\mu$m band is quite weak in the dark-QCC and 
the peak position of the 3.3\,$\mu$m
is difficult to determine (see Fig.~\ref{fig5}).  The feature
almost fades away in the 99\% \element[][13]{C} sample.  The change in the 
profile of the 3.3\,$\mu$m
seen in the filmy-QCC sample is not observed in the dark-QCC.  

%

\subsection{Peak shifts of the absorption bands}

 \begin{figure*} 
 \begin{center}
   \includegraphics[width=14cm]{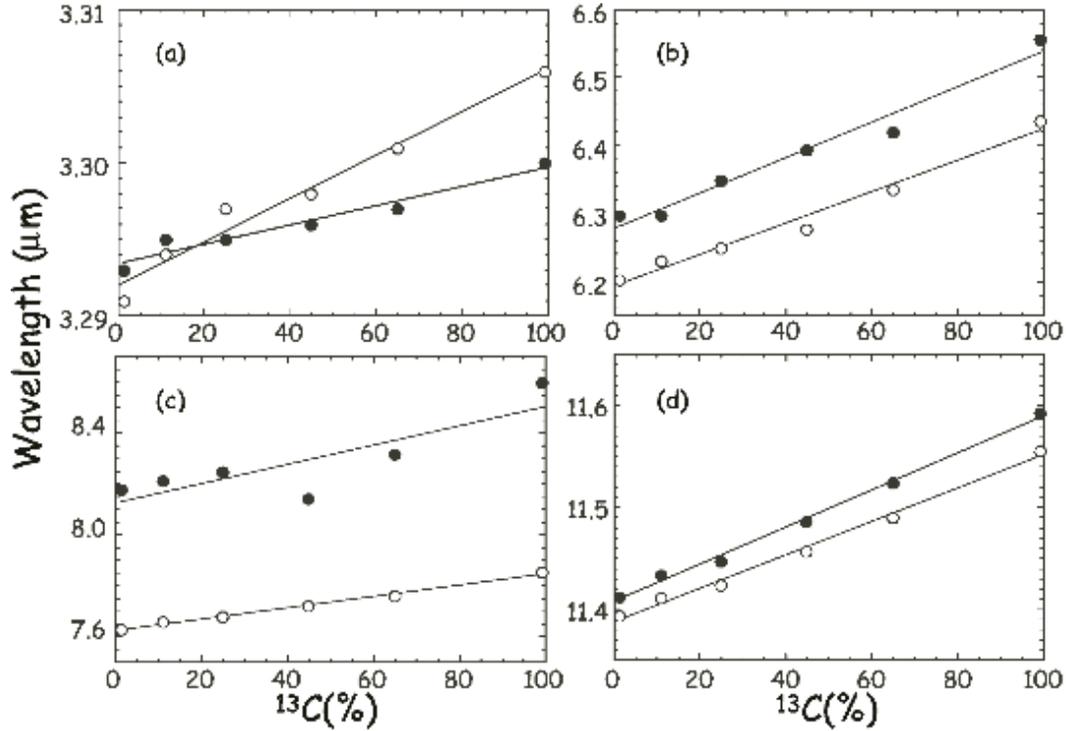}
\caption{Peak wavelength versus \element[][13]{C} fraction of the 
filmy-QCC 
(open circles) and the 
dark-QCC (filled circles). 
(a) Aromatic C$-$H stretching modes, (b) aromatic C$=$C 
vibration modes, (c) 7.6 and 8.2\,$\mu$m bands, and (d) C$-$H out-of-plane 
bending modes.  Least-squares fit lines are also plotted.
}
\label{fig6}
\end{center}
\end{figure*}

Fig.~\ref{fig6} plots the peak wavelength of each band against
the \element[][13]{C} fraction.  The change of the
peak wavelength is roughly linear with the \element[][13]{C} fraction
for all the bands plotted.  Linear lines obtained by least-squares
fits are also shown.
The slope of the fitted line is given in Table~\ref{tab1} together with
the shifts estimated from the calculation of 
\element[][12]C- and \element[][13]{C}-benzene
(Appendix~\ref{cal}).  The amount of the
shift in the 6.2\,$\mu$m
band is large and even larger than the estimate based on the simple 
calculation of benzene molecules.  
The shift of the 11.4\,$\mu$m band is also larger than the calculation.  
The shift in the 3.3\,$\mu$m band
of the filmy-QCC is in agreement with the calculation, while that of
the dark-QCC is smaller than the calculation.  It should be noted, however, 
that the 3.3\,$\mu$m band of dark-QCC
is weak and the peak position cannot be determined
accurately (Fig.~\ref{fig5}).
\begin{table*}
\caption[]{Peak shift of the absorption bands in the QCC and the UIR bands}
\begin{center}
\begin{tabular}[h]{ccccc} 
\hline \hline 
Peak wavelength & \multicolumn{2}{c}{Slope of the fitted line}
& Estimated shift between & Observed range of\\ of \element[][12]{C} QCC
& \multicolumn{2}{c}{($\mu$m per \element[][13]{C} fraction)}
& \element[][12]{C} and \element[][13]{C} benzene$^a$ & the 
peak wavelength$^b$\\
($\mu$m) & filmy-QCC & dark-QCC & ($\mu$m) & ($\mu$m) \\
\hline 
3.3  & $ 0.014 \pm 0.001 $ & \,\,$0.0063 \pm 0.0007 ^c$ & 0.013 & 3.288 -- 
3.297 
\\
6.2  & $0.231 \pm 0.018$ & $0.262 \pm 0.023$ & 0.14 & \qquad 6.20 -- 6.27 (6.30)\\
7.6  & $ 0.219 \pm 0.010$ & --- & --- & 7.72 -- 7.97$^d$\\
8.2  & --- & \,\,$0.379 \pm 0.124 ^c$ & --- &($\sim 8.22$) \\
11.4 & $0.163 \pm 0.007$ & $ 0.183 \pm 0.007$ & 0.04 & 11.20 -- 11.25\\
\hline
\end{tabular}
\end{center}
$a$ See Appendix~\ref{cal}.

$b$ The observed range between class $A$ and $B$ objects 
taken from Tokunaga et al. (\cite{tok91}), Peeters et al. (\cite{pee02}), 
and van Diedenhoven et al. (\cite{van03}).  The peak wavelengths of
class $C$ objects are indicated in parentheses (see text).

$c$ The slopes have large uncertainties because of the broad feature.

$d$ The peak wavelength range of the 7.8\,$\mu$m subcomponent 
(see text).
\label{tab1}
\end{table*}

The present experiment
shows no new features corresponding to \element[][13]{C}.  If the
vibration mode is local, there should appear peaks corresponding
to \element[][12]{C} and \element[][13]{C}.  The spectral resolution
is sufficiently high except for the 3.3\,$\mu$m band, which shows 
only small shifts, but any appreciable broadening of the feature that
indicates a combination of multiple peaks
is not seen with the \element[][13]{C} fraction.  
Thus the peak shift cannot be interpreted in
terms of a combination of the components corresponding to 
\element[][12]{C} and \element[][13]{C}.  The present results
indicate that the vibration modes in the QCC are not very localized, but
involve rather large molecular structures.  The peak shift is related to 
the number of carbon atoms involved in the
vibration mode.  The larger shift than the calculation
suggests that more than two carbon
atoms are associated with the 6.2\,$\mu$m band of the QCC.  The large
shift in the 11.4\,$\mu$m band also indicates that this feature originates
from CH bonds associated with large benzene structures if it is ascribed
to a CH out-of-plane bending mode.

The two kinds of the QCC have very different chemical compositions from each 
other.
The spectrum of the dark-QCC shows a continuum
component, which indicates the development of sp$^2$ carbon-carbon bonds.
The broad bump around 8.2\,$\mu$m and the strong
continuum are the characteristics of carbon-rich particles.
The filmy- and dark-QCC have absorption bands at similar wavelengths, 
but the bands seen in the dark-QCC always peak at longer wavelengths than
the filmy-QCC
except for the 3.3\,$\mu$m band with large \element[][13]{C} fractions
(Fig.~\ref{fig6}). 
The isotopic shift appears similar both in the filmy-QCC and  
dark-QCC except for the 3.3\,$\mu$m band.  
The 3.3\,$\mu$m band in the dark-QCC is rather weak and thus 
further examination is needed to understand the different behavior
of the 3.3\,$\mu$m band between the filmy- and dark-QCC.

\section{Discussion}

The infrared spectra of the QCC show several similarities to the observed UIR 
bands, 
but they are not exactly the same.  
The peak wavelength of the
3.3\,$\mu$m band of the f-QCC is in agreement with  
observations (Sakata et al. \cite{sak90}),
while the exact peak wavelengths
are not at the right positions for some other bands 
and there are differences in the band profiles.  The 11.4\,$\mu$m band in
the QCCs is obviously located at a longer wavelength than observed 
(11.23--11.25\,$\mu$m).  
Some UIR bands, such as the 6.2\,$\mu$m and 11.2\,$\mu$m bands,
are not symmetric, but skewed to longer wavelengths,
whereas most bands in the QCC seem to be more or less symmetric.
Therefore a direct comparison of the QCC spectra with   
observations is not very straightforward 
since the QCC is probably not the very
material that exists in interstellar space.  
However we
believe that the QCC contains many of the structural units corresponding
to the material in the interstellar medium because of the spectral
similarities of the QCC and the UIR bands.  
In the following we examine whether or not observed variations
of the UIR bands seen in post-AGB stars and PNe
can be accounted for by the isotopic shifts observed in the QCC.
We concentrate on the relative shifts in 
the band peaks of the QCC and the absolute peak wavelengths and band profiles
are not discussed.  
The results of the filmy-QCC are compared with the observations
for clarity because
the difference in the relative shifts between filmy- and dark-QCC is
small except for the 3.3\,$\mu$m band.

Peeters et al. (\cite{pee02}) have investigated the 6--9\,$\mu$m UIR bands
in various objects and found that there are at least three classes, designated
as $A$, $B$, and $C$, according to the peak wavelengths of the emission bands.  
Class $A$ is a major class and includes a wide
range of objects, such as \ion{H}{ii}
regions, non-isolated young stellar objects (YSOs), reflection nebulae, and
galaxies.   They do not show appreciable variations in the 6.2,
7.7, and 8.6\,$\mu$m band features.  Class $B$ consists of isolated YSOs,
post-AGB stars,
and PNe.  They exhibit quite a large variation in the peak wavelengths
of the 6--9\,$\mu$m features.
The peaks are all shifted to longer wavelengths compared to class
$A$ objects.  The observed
ranges of the peak wavelengths are listed in Table~\ref{tab1}.
The apparent variation in the 7.7\,$\mu$m complex comes partly from the
change in the relative strengths of the subcomponents.  The peak shift
in the 7.8\,$\mu$m subcomponent is well observed, while that in the
7.6\,$\mu$m subcomponent is less clear.  Table~\ref{tab1} indicates the
observed range of the peak wavelength of the 7.8\,$\mu$m subcomponent.
Class $C$ consists of only two objects (CRL\,2688 and IRAS\,13146$-$6243).  
They show a very 
different spectrum from the other two classes.  They have a band at 6.3\,$\mu$m
but do not show any features around 7.7 and 8.6\,$\mu$m.  Instead they show 
a broad
band at 8.22\,$\mu$m.  Hence class $C$ objects seem to have
different kinds of the band carriers from those in classes $A$ and $B$.
In the following discussion  
the peak wavelength of the 6.3\,$\mu$m band in class $C$ objects
is not included in the observed variation since it will probably not be
directly related to the isotopic effects.
Although there are some internal variations present, 
the class in the 6--9\,$\mu$m bands is
well defined and almost all class $B$ objects show the peaks shifted
to longer wavelengths 
in all the 6.2, 7.7, and 8.6\,$\mu$m bands.

Compared to the 6--9\,$\mu$m spectrum, the 3.3 and 11.2\,$\mu$m UIR bands
show small variations (Hony et al. \cite{hon01}), but the variations in
the peak position do exist 
(e.g. Nagata et al. \cite{nag88}; Tokunaga et al. \cite{tok88};
Roche et al. \cite{roc91}).  Tokunaga et al.
(\cite{tok91}) have suggested that there are two types of the 3.3\,$\mu$m band
from a small number of objects.
Type 1 is a major class, in which the peak is located at 3.289\,$\mu$m, while
type 2 has the peak at 3.296\,$\mu$m.  They also suggest that type 2 objects
have a narrower width than type 1.
van Diedenhoven et al. (\cite{van03}) have extended the
investigation of the 3.3 and 11.2\,$\mu$m
UIR bands on the same sample as in Peeters et al. (\cite{pee02}).  
They found that there are also two classes in each band, designated
as $A_{3.3}$, $B_{3.3}$, $A_{11.2}$, and $B_{11.2}$, respectively.
Class $A_{3.3}$ corresponds to type 1 and $B_{3.3}$ to type 2 in
Tokunaga et al. (\cite{tok91}).  $A_{3.3}$ and $A_{11.2}$ are major
classes, and $B_{3.3}$ and $B_{11.2}$ classes show the bands at
3.3\,$\mu$m and 11.2\,$\mu$m shifted to longer wavelengths compared
to $A_{3.3}$ and $A_{11.2}$, respectively. 
The observed ranges of the shifts in the 3.3 and 11.2\,$\mu$m bands
are also shown in Table~\ref{tab1}.  The shift in the 3.3\,$\mu$m band
is quite small and that in the 11.2\,$\mu$m band is modest.
In general objects classified as class $B$ in the 6--9\,$\mu$m
spectrum are also classified as $B_{3.3}$ and $B_{11.2}$,
but the correlation is not very good.  Objects classified
as class $B$ in the 6--9\,$\mu$m spectrum sometimes
do not show detectable shifts in
the 3.3 and 11.2\,$\mu$m bands (e.g. BD +30 3639) and vice versa
(e.g. He2$-$113).  
van Diedenhoven et al. (\cite{van03}) concluded that
the appearance of the 3.3 and 11.2\,$\mu$m band variations is not tightly
correlated with the UIR bands in 6--9\,$\mu$m.

The present results suggest that if \element[][12]{C}/\element
[][13]{C} $<10$ the isotopic shifts are detectable particularly
in the 6.2 and 7.7\,$\mu$m bands.  
A low \element[][12]{C}/\element[][13]{C} ratio of this range
is often suggested in post-AGB stars and PNe, and thus part of
the peak shifts in these objects could be attributed to the isotopic
effects.  
The shift in the 11.2\,$\mu$m
of a medium degree can also be expected, while the shift in the
3.3\,$\mu$m should be small.  This shift pattern is qualitatively
compatible with observations except for the 7.8\,$\mu$m band.  
The 7.8\,$\mu$m subcomponent shows a larger shift
in its peak wavelength compared to 
the QCC sample.  The 7.7\,$\mu$m complex exhibit complicated profile
variations and there may be other weak subcomponents present.
The observed range of
the 3.3\,$\mu$m, 6.2\,$\mu$m, and 11.2\,$\mu$m bands suggest that
the objects with the largest shifts should have
\element[][12]{C}/\element[][13]{C} $\sim 2$. 

\begin{figure*}
 \begin{center}
   \includegraphics[width=14cm]{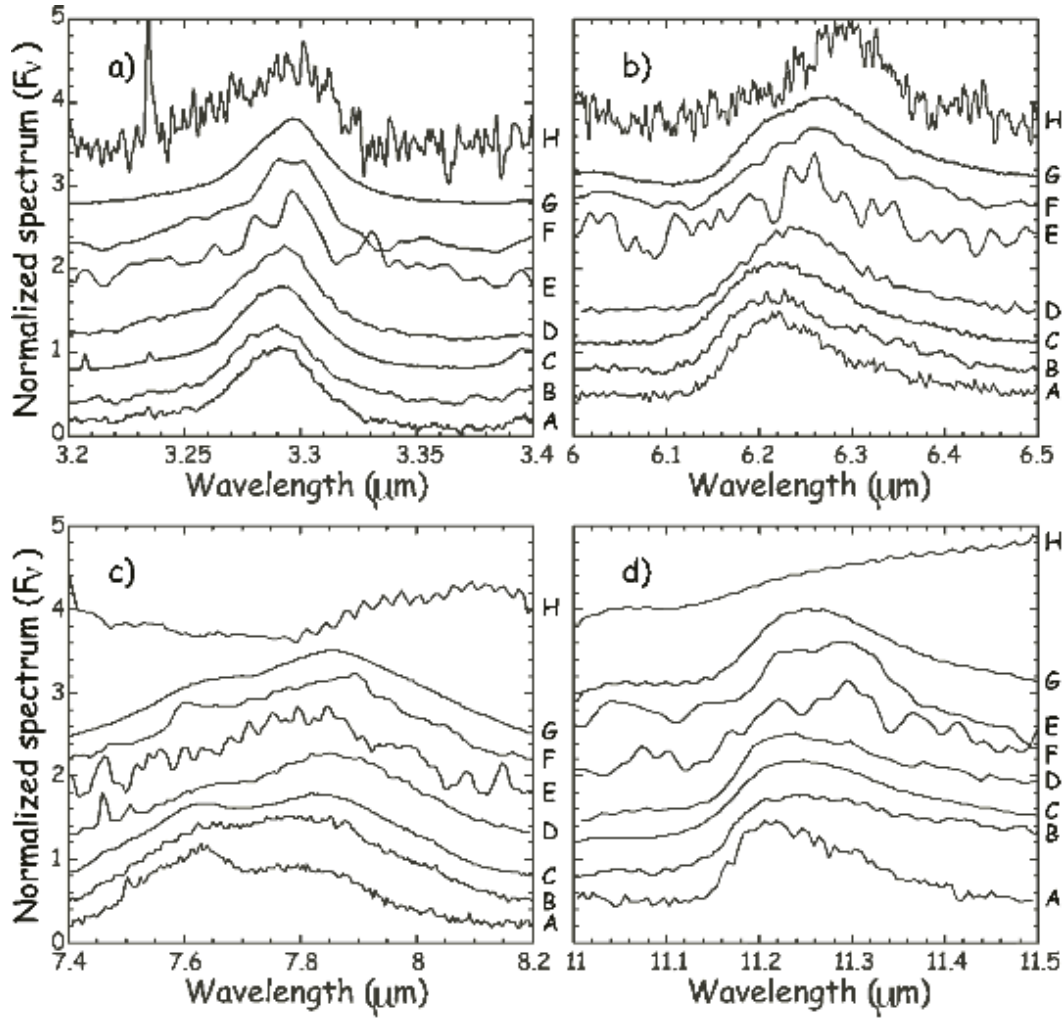}
\caption{UIR band spectra of the objects with the measured 
\element[][12]{C}/\element[][13]{C} ratio. a) the 3.3\,$\mu$m band, b)
6.2\,$\mu$m band, c) 7.8\,$\mu$m band, and d) 11.2\,$\mu$m band.
A: IRAS\,23133+6050, B: IRAS\,21282+5050, C: NGC\,7027, D: BD\,+30\,3639,
E: IC\,5117, F: HR\,4049, G: HD\,44179, and H: CRL\,2688. 
 The spectra are normalized after the local continua have been subtracted.
 Strong line emissions have been removed for clarity.
}
\label{fig7}
\end{center}
\end{figure*}

We have searched in literature for the \element[][12]{C}/\element[][13]{C} ratio 
of objects that had been observed by ISO/SWS.  
Table~\ref{tab2} lists the objects and the SWS observing modes.  SWS01
was the full grating scan mode and the spectral resolution depends on 
the scan speed.  SWS06 was the fixed-range  
grating scan mode
and provided the highest spectral resolution in the grating mode of SWS
(Leech et al. \cite{lee02}).
We used the highest spectral resolution data as much as possible if there
are several observations made for the same object.  IRAS\,23133+6050 is
included in the list as a reference of the class $A$ object.
The Off-Line Processing data of version 10.1 (OLP 10.1) were obtained from
the ISO Archival Data Center and reduced further by the Observers SWS 
Interactive Analysis
Package (OSIA) version 3.0.\footnote{The OSIA is a joint development of 
the SWS consortium. Contributing institutes are SRON, MPE, KUL and the 
ESA Astrophysics Division.}   The peak wavelength for each band is
estimated after the local continuum has been subtracted (see Fig.\ref{fig7}).
The peak position is slightly dependent on the assumed local continuum.
Strong Pf$\delta$ (3.29699\,$\mu$m) has been removed in the estimation
of the
peak wavelength of the 3.3\,$\mu$m band for IRAS\,23133+6050, NGC\,7027,
and BD\,+30\,3639.
Table~\ref{tab3} lists the objects with the derived isotope 
ratio and their observed peak wavelengths.  The peak wavelengths are
in agreement with previous investigations within the uncertainty 
(Nagata et al. \cite{nag88}; Tokunaga et al. \cite{tok91}; Peeters et al.
\cite{pee02}).  The objects are listed in the order of the peak wavelength
of the 6.2\,$\mu$m band.  Each band spectrum is normalized at the peak
after the continuum has been subtracted
and shown in Fig.\ref{fig7}.  

   \begin{table}
     \caption[]{Objects and the ISO/SWS observation mode}
     \begin{center}
     \begin{tabular}[h]{lccl}
     \hline
     \hline
     Name & Type & TDT$^a$ & Obs. mode$^b$ \\
     \hline
     \object{IRAS\,23133+6050}$^c$ & \ion{H}{ii} region & 56801906 & SWS01 (2)\\
     \object{IRAS\,21282+5050} & post-AGB & 15901777 & SWS01 (3) \\
     \object{NGC\,7027} & PN & 33800505 & SWS06 \\
     \object{BD\,+30\,3639} & PN & 86500540 & SWS01 (3) \\
     \object{IC\,5117} & PN & 36701824 & SWS01 (1) \\
     \object{HR\,4049} & post-AGB & 17100101 & SWS01 (2) \\
     \object{HD\,44179} & post-AGB & 70201801 & SWS01 (4) \\
     \object{CRL\,2688} & post-AGB & 33800604 & SWS06 \\
     \hline
     \end{tabular}
   \end{center}
   $a$ Target Dedicated Time (TDT) of the ISO observation.\\
   $b$ Number in brackets indicates the scanning speed of the SWS01 mode. \\
   $c$ Reference object of class $A$.\\
     \label{tab2}
   \end{table}

   \begin{table*}
     \caption[]{Peak wavelengths and carbon isotope ratios of the objects
     observed with ISO/SWS}
     \begin{center}
     \begin{tabular}[h]{lllllll}
     \hline
     \hline
     Object & \multicolumn{4}{c}{Peak wavelength ($\mu$m)}&
      \element[][12]{C}/\element[][13]{C}  & 
      Reference$^b$ \\
      & 3.3\,$\mu$m  & 6.2\,$\mu$m  & 7.8\,$\mu$m   &
      11.2\,$\mu$m   \\
      \hline
      IRAS\,23133+6050 & 3.289 & 6.219 & 7.795 & 11.218 & & \\
      IRAS\,21282+5050 & 3.290 & 6.219 & 7.817 & 11.245 & $>200$, $>32$ & 1, 2\\
      NGC\,7027  & 3.292 & 6.222 & 7.830 & 11.247 
      & $>31$, $>65$, $>25$, $>11$ & 3, 4, 5, 6 \\  
      BD\,+30 3639 & 3.292  & 6.240 & 7.850 &  11.237 & $\geq4$ & 7 \\
      IC\,5117 & 3.295: & 6.25: & 7.82: & 11.28: & 23, 14 & 6, 7 \\
      HR\,4049 & 3.296:$^c$ & 6.256 & 7.880: & 11.285 
      & $\sim 1.5^a$ & 8 \\
      HD\,44179 &  3.296 & 6.267 & 7.861 & 11.250 &
      $\geq 22$, $\geq 12.5$ & 3, 9\\      
      CRL\,2688 & 3.297 & 6.290 & \quad 
      -- & \quad --   & $>19$, $32^{+10}_{-7}$, 
      $>3$, $20^{+8}_{-6}$, 5& 3, 4, 5, 10, 11 \\ 
     \hline
     \end{tabular}
   \end{center}
   $a$  Intensity ratio of CO$_2$ bands (Cami \& Yamamura \cite{cam01})\\
   $b$  References\\
1: Likkel, et al. (\cite{lik88}); 2. Balser et al. (\cite{bal02});   
3: Bakker et al. (\cite{bak97}); 4: Kahane et al. (\cite{kah92}); 
5: Bachiller et al. (\cite{bac97}); 6: Josselin \& Bachiller (\cite{jos03});
7: Palla et al. (\cite{pal00});  8: Cami \& Yamamura (\cite{cam01}); 
9: Greaves \& Holland (\cite{gre97}); 10: Wannier \& Sahai (\cite{wan87});
11: Jaminet et al. (\cite{jam92}).\\
   $c$ Doubly peaked.\\
     \label{tab3}
   \end{table*}

Very few data of the carbon isotope abundance are available for post-AGB stars.
Observations of \element[][13]{C} isotopic lines of C$_2$, CN, and 
\element[+][]{CH} 
in several post-AGB stars
resulted in negative detection of molecular features of
\element[][13]{C} except for one object (Bakker et al. \cite{bak97, bak98}).
They thus provided only lower limits for the 
\element[][12]{C}/\element[][13]{C} ratio of $>10-20$.
The carbon isotope ratios in a few post-AGB stars and several
PNe have been obtained
mostly from radio observations of CO molecules.  
The intensity ratio of 
\element[][12]{CO} to \element[][13]{CO} emission
is thought to be a good measure for the isotope ratio if both
lines are optically thin since
the selective photodissociation of \element[][13]{CO} should
be counterbalanced by the charge exchange reaction (e.g. Mamon et 
al. \cite{mam88}).  However \element[][12]{CO} lines often become
optically thick, only lower limits of \element[][12]{C}/\element[][13]{C}
being estimated.
It should also be noted that observations of CO probe the entire
region of 
the circumstellar envelope.  The post-AGB is a transient phase and a rapid change
of the abundance is expected to occur in their envelopes. 
Thus the isotope ratio in the vicinity of the star might
be different from that inferred from the CO intensity ratio.

IRAS\,21282+5050 is a post-AGB star and classified as class $A$ in the
6--9\,$\mu$m spectrum and $A_{3.3}$.  The 3.3\,$\mu$m band peaks around
3.288--3.290\,$\mu$m
(e.g. Nagata et al. \cite{nag88}).  Likkel et al. (\cite{lik88}) obtained
that the intensity ratio
of \element[][12]{CO}/\element[][13]{CO} ($J$=1--0) is about 100
and inferred \element[][12]{C}/\element[][13]{O} $\sim 200$,
taking account of the optical thickness effect of \element[][12]{CO}.
Based on the large gradient velocity model analysis,
Balser et al. (\cite{bal02}) obtained 
\element[][12]{C}/\element[][13]{C} $>32$
from observations of CO $J$=2--1 and $J$=3--2 transitions.
The observed peak wavelengths are
compatible with the large \element[][12]{C}/\element[][13]{C} ratio.

NGC7027 is a bright young PN.  It shows the class $A$ 6.2\,$\mu$m band
feature, while the 7.7 and 11.2\,$\mu$m bands are slightly shifted
to longer wavelengths relative to class $A$.  Also the 7.8\,$\mu$m component
is stronger than the 7.6\,$\mu$m component, indicating a signature of the
class $B$ object.
The 3.3\,$\mu$m band peak appears at the normal (class $A$) wavelength.
This object has been observed by several groups in \element[][13]{CO} 
molecular lines.   Because of the optical depth effect of 
\element[][12]{CO} the CO line intensity ratio
should be taken as lower limits for the isotope abundance ratio
of \element[][12]{C}/\element[][13]{C}.  
They range from 65 (CO($J$=1--0); Kahane et al.
\cite{kah92}) to a recent value of 11 (CO($J$=2--1); Josselin \& Bachiller
\cite{jos03}), suggesting that 
the \element[][12]{C}/\element[][13]{C} ratio in the CO
envelope of NGC7027 is probably not less than 10.  Except for the
7.7\,$\mu$m band appearance
the suggested range of \element[][12]{C}/\element[][13]{C} is 
compatible with the class $A$ classification.

BD\,+30\,3639 is a young PN and classified as class $B$ in the 6--9\,$\mu$m
spectrum.  Palla et al. (\cite{pal00}) reported a tentative
detection of \element[][13]{CO} ($J=2-1$) and obtain a lower limit
of \element[][12]{C}/\element[][13]{C} as $\geq 4$.  This object shows
medium size band shifts, which can be accounted for by the isotopic shifts with
\element[][12]{C}/\element[][13]{C} $\sim 5$.  The corresponding
shift in the 3.3\,$\mu$m would be only $\sim$ 0.03\,$\mu$m, which is also
compatible with the observed peak position.  The 11.2\,$\mu$m band shows a
large tail in the longer wavelength side.

Assuming that CO lines are optically thin,
Palla et al. (\cite{pal00}) derived the 
\element[][12]{C}/\element[][13]{C} of 14 for IC5117, while
Josselin \& Bachiller (\cite{jos03}) indicated the ratio of 23 in this object. 
IC5117 was not included in the sample of Peeters et al. (\cite{pee02}).
The SWS spectrum is noisy and the peak wavelengths cannot
be determined accurately.  However it clearly indicates the characteristics
of the class $B$ spectrum.  The 7.7\,$\mu$m band complex is
dominated by the 7.8\,$\mu$m component and the 3.3 and 6.2\,$\mu$m bands
peak at longer wavelengths than class $A$.  The indicated 
\element[][12]{C}/\element[][13]{C} ratio seems to be too small to account 
for the observed spectrum by the isotopic effects, although a qualitative
comparison is difficult because of large uncertainties in the peak
wavelengths.

HR4049 is a very metal-poor post-AGB star and it is difficult to estimate
its carbon isotope ratio because of very few metallic lines detected
in its spectrum.  Recently
Cami \& Yamamura (\cite{cam01}) have investigated circumstellar CO$_2$
features in the infrared region and suggested that
this star has a very peculiar oxygen isotope ratio, such that \element
[][17]{O} and \element[][18]{O} are quite enhanced.  They have detected
\element[][13]{C}\element[][16]{O}$_2$, 
\element[][16]{O}\element[][13]{C}\element[][18]{O},
and \element[][16]{O}\element[][13]{C}\element[][17]{C} lines, and the
intensity ratios of the \element[][13]{C} to \element[][12]{C} lines were 
all about 0.6--0.7.  Although detailed
analysis is required to derive a reliable ratio, this implies
that HR4049 is quite rich also in \element[][13]{C}.  
The positive correlation of \element[][12]{C}/\element[][13]{C} with 
\element[][16]{O}/\element[][17]{O} seen in several environments 
(Wannier \& Sahai \cite{wan87}) also suggests a low 
\element[][12]{C}/\element[][13]{C} ratio. 
CO$_2$ molecular emission
originates from the region near the star and thus the derived ratio should
correspond to the region in which the UIR bands are also emitted.  
The observed
UIR bands in HR4049 are all shifted to longer wavelengths relative
to normal class $A$ positions, which can be accounted for by isotopic 
shifts with
\element[][12]{C}/\element[][13]{C} $\sim 3$.  
The small \element[][12]{C}/\element[][13]{C} ratio suggested by the CO$_2$ 
line emissions is compatible with the observed shifts.

HD44179 is a well-studied post-AGB star and very bright 
in the infrared. 
Greaves \& Holland (\cite{gre97}) made observations of CO 
($J$=2--1) and obtained the peak intensity ratio
of $2.2 \pm 0.2$ for \element[][12]{CO}/\element[][13]{CO}.  The central part of 
the \element[][12]{CO} profile appears
optically thick and from the wing of the profile they estimate 
\element[][12]{CO}/\element[][13]{CO}$\geq
12.5$.  The negative detection of \element[+][13]{CH} suggests 
\element[][12]{C}/\element[][13]{C} $\geq 22$ (Bakker et al. \cite{bak97}).
The 6--9\,$\mu$m spectrum of HD44179 belongs to
class $B$ and shows one of the largest shifts in the peak wavelengths
among the class.  HD44179 is also 
the only one object that shows two subcomponents in the 6.2\,$\mu$m
clearly.
The 3.3\,$\mu$m band of HD44179 is of type 2 and peaks at 3.296\,$\mu$m.  
The 11.2\,$\mu$m band is also shifted to 11.25\,$\mu$m.
These shifts can be accounted for consistently by the isotopic effects
if \element[][12]{C}/\element[][13]{C} is $\sim 2$.
This required
ratio is obviously smaller than the one indicated by CO and
\element[+][13]{CH} observations.  A complicated torus-bipolar cone
structure resulted from a transition
from oxygen-rich to carbon-rich envelope has been suggested
in HD44179 (Waters et al. \cite{wat98}; Men'shchikov et al. \cite{men02}), 
indicating that 
the carbon isotope ratio may also spatially vary.  In fact
Kerr et al. (\cite{ker99}) have indicated that the 3.3\,$\mu$m
band peak changes with positions within the nebula
and that the longer wavelength peak appears near the star and the interface
region of the biconical structure.  The \element[][13]{CO} emission
with a similar strength to \element[][12]{CO} has also been detected
for the 4.6\,$\mu$m bands in the nebular region (Waters et al. \cite{wat98}).
Two components seen in the 6.2\,$\mu$m band in HD44179 may then
be ascribed to the band carriers with a small \element[][13]{C} isotope 
abundance in the outer part and to those with a low
\element[][12]{C}/\element[][13]{C} formed in the region
close to the star.  Unless there is an appreciable increase in the
\element[][13]{C} abundance in the vicinity of the star, however, the 
observed peak shifts in HD44179 cannot be accounted for by isotopic
effects alone.
High-spatial resolution observations in other UIR
bands are interesting to investigate possible spatial variations 
in their peaks (Song et al. \cite{son03}).

CRL\,2688 is a carbon-rich post-AGB star which shows a peculiar
6--9\,$\mu$m spectrum (class $C$).  The integrated intensity ratio of 
\element[][12]{CO} to \element[][13]{CO} ($J$=2--1)
is about 3--4 (Wannier \& Sahai \cite{wan87}; Bachiller et al.
\cite{bac97}).  The \element[][12]{CO} line is optically thick and
thus this gives a lower limit.  Wannier \& Sahai (\cite{wan87})
derived the CO/\element[][13]{CO} ratio to be 20 from model analysis on 
CO ($J$=2--1) observations and Kahane et al. (\cite{kah92}) obtained
the ratio of $32^{+10}_{-7}$ from observations of CS.
Jaminet  et al. (\cite{jam92}) have indicated, on the other hand,
that the fast wind component has \element[][12]{C}/\element[][13]{C} of 5
from the wing profile of CO ($J$=3--2) emissions.
Bakker et al. (\cite{bak97}) gave a lower limit of 19 based on the
negative detection of circumstellar molecules with \element[][13]{C}.  
Goto et al. (\cite{got02}) indicated that the 3.3\,$\mu$m band profile
does not show any spatial variations, suggesting that the extended component
of the band emission is the scattered light of the emission from the 
central region.  There is no evidence so far for the spatial variation
in the band features within the nebula.
Since this
object may have a different type of the band carriers, 
the peak position of the 6.2\,$\mu$m band
(in fact it is located at 6.29\,$\mu$m) should be discussed separately from
other objects.  The 8.2\,$\mu$m band seen in the
dark-QCC may correspond to the band at 8.22\,$\mu$m detected
in class C objects.  It should be noted that 
the dark-QCC of the natural isotope abundance
also has a band at 6.3\,$\mu$m, 
shifted to a longer wavelength compared to the filmy-QCC.
The 3.3\,$\mu$m band is seen at the class $B$ position and
the 11.2\,$\mu$m is very weak, if present.  
Differences in the type of the band carriers themselves may account for
the observed shift in the 6.2\,$\mu$m band as suggested by the difference
between filmy- and dark-QCC spectra.  

As described above, carbon isotope ratios obtained from CO observations
of post-AGB stars and PNe are upper limits in most cases 
or show large scatter
because of uncertainties in the observed intensities of faint 
\element[][13]{CO} emissions (see Josselin \& Bachiller \cite{jos03})
and in the correction for the optical depth effects of 
\element[][12]{CO} lines.  In addition very few objects in the
post-AGB phase with the UIR bands have measured 
\element[][12]{C}/\element[][13]{C} data, including upper limits.
Thus it is difficult at present to directly compare
the present results with observations for individual objects.  
The overall shift pattern of the UIR band peaks is 
qualitatively in agreement with the isotopic shift pattern except for
the 7.7\,$\mu$m band complex.  However,
Table~\ref{tab2} does not indicate a quantitatively clear correlation
between \element[][12]{C}/\element[][13]{C} and the peak wavelengths and
the suggested \element[][13]{C} abundance in individual objects seems 
to be slightly
too small to account for the observed shifts in terms of the isotopic effects.  
Observations also indicate that the correlation among the peak wavelengths
of the 6--9\,$\mu$m bands 
is good, but the peak shifts in the 6--9\,$\mu$m bands
are not tightly correlated with those in
the 3.3 and 11.2\,$\mu$m bands.
The UIR band emission is thought to come from the vicinity of the
central star, while the CO emission probes the entire region
of the circumstellar envelope.  
Part of the lack of clear correlations and the poor quantitative agreement
between the isotope ratio and the peak wavelengths may be 
attributed to possible spatial variations in the isotope abundance
within the object.  If
the 3.3 and 11.2\,$\mu$m bands come from 
regions different (either closer to or farther away from the source)
from that of the 6--9\,$\mu$m bands, the poor correlation of the
peak wavelengths among the 3.3, 6--9, and 11.2\,$\mu$m bands may
also be attributed to the spatial variation in the
isotope ratio with which each band carrier was formed
in the envelope. 

Most PAHs have band peaks
longer than 6.3\,$\mu$m for a C$=$C vibration contrary to the filmy-QCC and
it has been suggested that the substitution
of a carbon atom by a hetero-atom, such as nitrogen, oxygen, or silicon,
will
shift the band to shorter wavelengths as observed in class $A$ objects
(Peeters et al. \cite{pee02}).  
Then the observed variations
are interpreted in terms of the relative abundance of pure-carbon PAHs
and substituted PAHs.  The hetero-atom substitution, on the other hand, 
does not make systematic shifts in other bands.  The isotopic
substitution of \element[][13]{C} has similar effects
on the peak wavelength for the 6.2$\mu$m band but 
in the opposite direction.  In addition it makes
systematic shifts in the peak wavelengths of other bands and 
qualitatively accounts
for the observed variations correlated with the 6.2\,$\mu$m band.  
The peak wavelength of the C$=$C stretching vibration band decreases
with the size of PAHs, while the increase of
the molecular size does not make systematic shifts in the 7.7\,$\mu$m
band (Peeters et al. \cite{pee02}).
Hetero-atom substitutions and size variations
are likely to occur in interstellar medium
and the observed variations could be a summation of various processes.
The present results indicate that the isotopic effects can  
contribute to the observed variations for objects with
low \element[][12]{C}/\element[][13]{C} ratios.
It should also be emphasized that the peak shifts of the UIR bands 
are observed mostly in post-AGB stars and PNe, which are the objects
where small \element[][12]{C}/\element[][13]{C} ratios are expected. 

It has been suggested that
deuterated PAHs (PADs) can be efficiently formed 
in dense molecular clouds because the large difference in the zero-point 
energy between H and D containing species leads to preferential formation
(fractionation) of deuterated species
at low temperatures (e.g. Tielens \cite{tie97}; Sandford et al. \cite{san00}).
For the carbon isotope case, the zero-point energy
difference is not large and the temperature in the circumstellar envelope
is much higher then the zero-point energy.  Hence isotopic
fractionation should not be significant. 
Instead a rather large \element[][13]{C} abundance is expected to
exist in certain environments and  
carbonaceous dust that is formed in such environments
should show isotopic shifts in its band features.
Shifts in the band
peaks of PADs from those of PAHs are quite large because of the 
large mass difference in
H and D.  Contrary to deuterated species,
expected shifts for \element[][13]{C} carbonaceous 
material are small but should still be in the detectable range
as shown by the present results.

\section{Summary}
We synthesized the QCC, a laboratory analogue of carbonaceous
dust, with various 
\element[][13]{C} 
fractions and measured the isotopic shifts in the peak wavelengths
of the infrared bands.
They all shift to longer wavelengths approximately linearly with
the \element[][13]{C} fraction.  The fact that no separate
peaks originating from \element[][13]{C}-QCC appear indicates
that the vibration modes of
infrared bands associated with carbon atoms in the QCC
should not be very localized, but that they 
stem from rather large structures containing several carbon atoms.
The shifts in the 6.2 and 7.8\,$\mu$m
are quite large ($\Delta \lambda > 0.2$\,$\mu$m per \element[][13]{C}
fraction) and that in the 11.2\,$\mu$m is modest 
($\sim 0.16$\,$\mu$m), while the shift in the 3.3\,$\mu$m band is
small ($<0.02 \mu$m). 
The small shift in the 3.3\,$\mu$m band is in agreement with
a simple calculation of benzene molecules and those in the 6.2 and
11.2\,$\mu$m seem to be larger, also suggesting that a larger number of
carbon atoms associated with these bands than benzene.
The isotopic shifts obtained in the present experiment are qualitatively
in agreement with the shift pattern observed 
in the peak wavelengths of the
UIR bands except for the 7.7\,$\mu$m band complex.  
However the observed variations seem to be larger than those
inferred from the \element[][12]{C}/\element[][13]{C} ratio
and the quantitative agreement
between the carbon isotope ratio and the peak wavelengths is
not good for individual objects.  This may be attributed partly to 
large uncertainties
in the isotope ratios derived from observations, to
possible spatial variations in the isotope abundance in
the envelope, and to combinations with other effects, such as
hetero-atom substitutions. 
The present results indicate that the isotopic
shifts in the peak wavelengths of the UIR bands should be detectable
in objects of low \element[][12]{C}/\element[][13]{C} ratios
and part of the observed variations in the band
peaks can be attributed to the isotopic effects.

\begin{acknowledgements}
This work is based in part on observations with ISO, an ESA project with 
instruments funded by
ESA Member States (especially the PI countries: France, Germany, the
Netherlands and the United Kingdom) and with the participation of ISAS and
NASA.   Part of this work was supported by Grant-in-Aids for Scientific Research
from Japan Society for the Promotion of 
Science.
\end{acknowledgements}

\appendix
\section{Calculation of isotopic peak shifts of benzene molecule}\label{cal}
In this appendix, we estimate isotopic peak shifts in the vibration
modes of a benzene molecule
to compare with the isotopic shifts in the corresponding vibration modes
of the QCC.  Benzene is the simplest PAH and the results should be used
as a 0-th order reference in comparison with the isotopic shifts in the QCC.

The $\vec{GF}$ matrix method is used to calculate isotopic shifts of
a benzene molecule (Wilson et al. \cite{wil55}).
The $\vec{G}$ matrix represents the molecular geometry and the $\vec{F}$ 
matrix consists of the force 
constants.  The eigenvalue of the $\vec{GF}$ matrix is given by 
$\lambda=2\pi^2\nu^2$, where $\nu$ is the frequency of the vibration mode.  
In the calculation of the isotopic shift, only the 
values of the $\vec{G}$ matrix elements change due to the isotopic substitution 
and 
the $\vec{F}$ matrix elements remain unchanged if
all carbon atoms are substituted by \element[][13]C. 
We use the values of the force 
constants of benzene given by Crawford \& Miller (\cite{cra46, cra49}).
The vibrational 
modes of infrared active fundamentals in benzene consist of one 
$A_\mathrm{2u}$ 
(C-H out-of-plane bending) and three $E_\mathrm{1u}$ (three kinds 
of in-plane vibrations) modes.  The single 
$\vec{G}$ matrix element of $A_{\rm 2u}$ is written by
\begin{equation}
\vec{G}(A_\mathrm{2u}) =  (\mu_\mathrm{H} + \mu_\mathrm{C})~ \sigma^2 
\label{a1}
\end{equation}
and those of $E_\mathrm{1u}$ are given by 
\begin{eqnarray}
\lefteqn{\vec{G}(E_\mathrm{1u})} \nonumber \\
& = \left( \begin{array}{ccc}
\mu_\mathrm{H}+\mu_\mathrm{C} & -\frac{\sqrt{3}}{2}\mu_\mathrm{C} & 
\frac{3}{4}\tau\mu_\mathrm{C} \\  & \frac{3}{2}\mu_\mathrm{C} & 
-\frac{\sqrt{3}}{4}(2\sigma+3\tau)\mu_\mathrm{C} \\
 & & \sigma^2\mu_\mathrm{H}
 +(\sigma^2+\frac{3}{2}\sigma\tau+\frac{9}{8}\tau^2)\mu_\mathrm{C} \\ 
\end{array} 
\right), \nonumber \\
\label{a2}
\end{eqnarray}
where only the upper right elements are shown for the symmetric 
$\vec{G}$ matrix. 
The parameters $\sigma$ and $\tau$ indicate the reciprocals of the 
C$-$H and C$-$C bond 
lengths, while $\mu_\mathrm{H}$ and $\mu_\mathrm{C}$ are the 
reciprocal masses of H and C, respectively.  The bond angles are all set as 
120$^{\circ}$. 
From Eqs.(\ref{a1}) and (\ref{a2}) and the $\vec{F}$ matrix we calculate the 
eigenvalues of the $\vec{GF}$ matrix and obtain the wavelength ratios
of each vibrational mode between \element[][12]{C}-benzene and  
\element[][13]{C}-benzene.  Then the wavelengths of \element[][12]{C}-benzene 
are scaled to the measured values.  The results are listed in Table~\ref{ta}.
Because the wavelength of each mode is
different from that in the QCC, we scale the wavelengths of 
\element[][12]{C}-benzene
to the bands measured in the 
\element[][12]{C}-QCC and estimate the isotopic shifts 
(Table~\ref{tab1}).  The $\lambda_{20}$ mode is 
scaled to the 3.3\,$\mu$m band and the $\lambda_{19}$ mode
to the 6.2\,$\mu$m 
band.  The band at 11.4\,$\mu$m in the QCC is assigned to a solo or 
duo 
C$-$H out-of-plane bending, which does not have a corresponding mode in 
benzene. 
 We estimate the isotopic shift by simply scaling the $\lambda_{11}$ mode to 
11.4\,$\mu$m because it is also a C$-$H out-of-plane mode of benzene.  

 \begin{table}
   \caption[]{Wavelength of the vibration modes of \element[][12]{C}-  
and \element[][13]{C}-benzene}
   \begin{center}
\begin{tabular}{ccc} \\ \hline 
mode &  \element[][12]{C}-benzene & \element[][13]{C}-benzene\\
     & ($\mu$m)     & ($\mu$m)      \\
\hline 
   $\lambda_{20}$ & 3.245  & 3.257 \\
   $\lambda_{19}$ & 6.812           & 6.964  \\
   $\lambda_{18}$ & 9.524           & 9.709  \\
   $\lambda_{11}$ & 14.90             &   14.95  \\
\hline

\end{tabular}
\end{center}
\label{ta}
\end{table}
%

\end{document}